IRJEMS International Research Journal of Economics and Management Studies
Published by Eternal Scientific Publications
ISSN: 2583 – 5238 / Volume 2 Issue 2 April 2023 / Pg. No: 87-92
Paper Id: IRJEMS-V2I2P110, Doi: 10.56472/25835238/IRJEMS-V2I2P110*Original Article*

# Why Students Trade? The Analysis of Young Investors Behavior

[1]Jones Pontoh
[1]Faculty of Economic and Business, Manado State University, Manado, Indonesia.Received Date: 09 April 2023      Revised Date: 19 April 2023      Accepted Date: 23 April 2023      Published Date: 29 April 2023***Abstract:*** *Interestingly the numbers of young traders in Jakarta Stock Exchange had been increasing in recent years. Even in the middle of the global crisis caused by covid19 pandemic, in december 2021 according to KSEI, Individual investors were dominated by young investors. Data presented by KSEI showed that 60% of the investors listed in Indonesian Stock Exchange were young investors (under age 30). Other data shows that 28% of the investors listed were shockingly students. It was interesting to study the behavior of young and "Rookie" investors at "Gallery Bursa Efek" of Manado State University. Basicaly they varied in how to make decision to trade on the stock exchange. The problems were discussed by qualitative approach. Descriptive analysis was conducted prior to interviews. Data will be collected through data observation techniques and interviews. The study succeded in investigating the investment behavior of young or Rookie investors at Manado State University in accordance to investment decision making and the perception of behavioral control.  The preception of behavioral control greatly influenced investors decision making. Students were greatly influenced by lecturer, friends and more experienced investors. The results of the interview provide information that before they determine their behavior, first do stock analysis, both technical and fundamental analysis. These facts shows that students investors were well literated.*

***Keywords:***  *Indonesia Stock Exchange, Investment, Capital Market, Investor Behavior.*## I. INTRODUCTION

The capital market is one of the most important sectors in Indonesian business development. This sector is still dominated by large investor investors  both domestically and abroad. The government sees this as a challenge for young people to be able to get involved and become the backbone of Indonesia's financial revival from the capital market sector. From indonesian investor data we saw that there are incrising number investor mostly on young investors. Data shows that under 30 age category had 60% of total investor, and 28% of the investors listed were student. Furthermore, 52% of the investor had income range from 10 Million to a hundred million rupiahs. These facts showed that there were tendecies among young investors to participate in the market.

Indonesian government had been struling to increase the number of investors in the past 10 years. One of the government's efforts was to encourage investment management companies to come to universities in order to provide the facilities and knowledge needed by young people so that they can be actively involved in the capital market. The increase in the number of investors will have a direct impact on the investment company and indirectly contribute to the impact on the development  of the capital market world I ndonesia.

Since a few years ago, the Faculty of Economics and Business in Manado State University (UNIMA) had been inisiating and improving the "Galery Bursa Efek" which was a place to trade provided by the University in cooperation with a particular investment management companies in Manado. Along the way, this small stock market experiences ups and downs every year. From interviews with the managers of the corner of the exchange, it is known that student participation in the trading activities at "Gallery Bursa efek" were still seasonal trends. At some times there were groups of students who were very interested in working in the world of capital markets. Even in some local competitions, UNIMA students managed to be the best in quantity and quality of exchange members. But at other times the number of active students is very small.

In interviews with some young investors, there were interesting things found. First was that some students still thought that membership in the corner of the exchange was just a requirement to pass some courses. They didn't see the need to be more actively involved because membership is just a prerequisite for graduation. This resulted in a lack of balance in the number of exchange members with the total number of transactions in the corner of the exchange.

The second problem encountered is that some student groups   felt safer if they made transactions using references from transactions that have been made by larger stock exchange investors.  They still have doubts about their own analysis.  Various

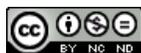This is an open access article under the CC BY-NC-ND license (https://creativecommons.org/licenses/by-nc-nd/2.0/)

behaviors of new investors greatly influenced these young investors

The above problems directly or indirectly have an impact on the interest and activeness of members of the corner of the exchange. This study aims to analyze the problems that exist related to student transactions of exchange members, background, motivation for student perception in transacting. Based on the description of the problem above, the focus in this study were, first, what were young investor investment/ transaction considerations?   Second, How did young investor trading behavior?

## II. MATERIALS AND METHODS

### A) Capital Market Theory

The capital market was essentially a market, not much different from the traditional market that we had known so far. There were merchants, buyers, and for sure, bargaining prices. Capital Market was defined as activities related to Public Offering and trading of Securities. Public Companies related to the Securities issued by them (Investment Managers with Mutual Funds issued by them), and institutions and professions related to Securities. From the definition above, it is clear that the Capital Market includes all activities and also actors related to Securities. Securities were debt recognition letters, commercial papers, stocks, bonds, proof of debt, Units of Participation in collective investment contracts, futures contracts on Securities, and any derivatives of Securities. The capital market provided various investment alternatives for investors in addition to the investment alternatives that we have known before such as saving in banks, buying gold, land, insurance and so on. The Capital Market acted as a liaison between investors and companies or government institutions through long-term financial instriumen trading such as Bonds, Stocks and others.

### B) The Role of the Capital Market

If there was no traditional market, then we could imagine how difficult it would be for us to find daily necessities such as spinach vegetables, for example. If we need spinach vegetables then we have to tell our neighbors whether anyone has spinach plants and we can buy them. That's just one need even though our needs for 1 day are very much. With the market, we can obtain these needs in an easy way and the time needed is also relatively short.

Likewise, the Capital Market was the most efficient place for allocating funds because with the capital market, an investor who had excess funds does not need to go around to ask whether the company was in need of funds in every company he meets, or vice versa the company did not need to spend unnecesary resources to find investors. Before the capital market was invented, if we want to take advantage of the excess funds, we bought gold, land, or started our own business. If we needed funds and want to sell the land, it was not so easy, because there were not necessarily people who interested in the land we have. Meanwhile, if we wanted to set up our own business, we need competence, experience in the new business line not to mention the big fund we have to raise.

With the capital market, we had other alternatives in investing that provide profits with a certain amount of risk. The market generated profit, by buying bonds or shares of a company then we will get interest that was greater than deposits (for bonds) or profit sharing (for stocks). However, there was risks that we must bear, such as the risk of the company's inability to pay debts or the bankruptcy of the company so that the shares we have will be lost. One of the requirements for a company to enter the capital market is to have good performance and be able to publish its financial statements every certain period regulated in the capital market. Thus, before the investor enters the capital market, he can first learn which company's shares he will choose.  The positive side for companies that enter the capital market is that the company will be motivated to always improve its performance so that the company will be more professional, efficient and profit-oriented, while because part of the company's ownership is in the hands of the general public, the company must be transparent. With transparency, the company will be encouraged to be healthier. With the capital market, it would be easier for companies to get funds to expand their business. This will lead to increased job opportunities for the community, which will further encourage the national economy to become more advanced. On the other hand, with the increase in company profits, tax gains for the government will also increase.

### C) Perception of Behavioral Control

Perception of behavioral control or also known as behavioral control was a person's feelings regarding the ease or difficulty of realizing a certain behavior, (Ajzen, 2005). Based on the concept of perception of behavior control, it is expected to reduce the relationship of intention to individual treatment, therefore only when the perception of individual behavior control is also strong can strong intentions give rise to behavior. Trust in the supported thing and obstacles to the implementation of the inhibitory treatment based on about the previous behavioral experience of the person, the person's understanding of the behavior, and obtained by observing his knowledge and other familiarity. Individuals and various other factors known to individuals can add or decrease an individual's perception of the degree of distress in carrying out behavior In the theory of planned behavior in particular. Behavioral control is defined as a person's perception of the ease or difficulty in carrying out a behavior, The perception of behavioral control can determine in terms of a person's beliefs in relation to supporting or obstacles in carrying out a behavior (control beliefs). In particular, in the material of behavior planning, this perception of behavioral control is as a theory of planned behavior about the difficulty and ease of carrying out behavior. The perception of behavioral control is determined by the combination of individual beliefs in relation to the supporting factors of the inhibitor in carrying out a behavior (control beliefs), over the strength of the individual's feelings in relation to each of those supporting or inhibiting factors.

Specifically, in the theory of planned behavior, the theory of perception of behavioral control can be interpreted in the behavior of the individual about the difficulty or not in perceiving his behavior. The perception of behavioral control can determine from the combination of one's beliefs in relation to the thing that supports and vice versa in providing a behavior (control belief), by paying attention to the feelings of the thing in relation to each of those supporting or inhibiting factors. (Perceived control of power). The more people perceive the difficulty or not and the more limited the false intensive in carrying out a behavior, the easier it is for a person to perceive the behavior. to do. On the contrary, fewer and fewer people feel some factors that support and hinder to carry out a behavior, so that the individual can find himself difficult to perform, The perception of the control of the investor's behavior is reflected in how much confidence, the capital that supports, and the use of technology.

*D) Methods*

This research used a qualitative approach and the methods were more fundamental to phenomena that prioritize evaluation. The qualitative descriptive method, according to Sugiono (2013) was that research can be carried out to determine the value of a independent variable, or more variables, without comparing or associating it with other variables. This research described interviews with research subjects studied in detail to convey a clear picture of the behavior of young investors in making investment decisions in the capital market. The population in this study was young investors who trade in the Manado State University "Galery Bursa Efek". Interviews were used as a technique of collecting data at the time when the researcher wants to conduct a preliminary survey to determine the problem being investigated with the aim of wanting to obtain more detailed information about the respondents. Preparation The interviewer who will prepare for an interview, needs to make some questions to be asked, the order of the questions, how important the questions are, the timing of the interview, and how to formulate the questions. Observation wass a method of collecting qualitative data directly through observation at the research site in making stock investment decisions. observation is the systematic observation and recording of the elements that appear in a symptom in the object of study according to Widoyoko (2014). Documentation according to Sugiyono. (2015) is a way to obtain data and information in the form of books, documents, document numbers, and photos in the form of reports and information to support research. To validate the interview results triangulation method was used. Triagulation in data collection was a data collection technique that combines various data collection techniques and existing data sources. if the researcher collects data by triangulation, then the researcher was actually collecting data that at the same time tests credibility. Sugiyono (2021)

*E) Research Instrument*

Data collection in qualitative research using various research methods such as observation, interviews, documentation and triangulation, and requires tools as instruments. The instruments used by researchers were smartphone as a voice recorder and taking pictures and digital interview forms in documenting key information from informants.

*F) Data Analysis Techniques*

   **a. Data Reduction**

   Data reduction means summarizing focusing on what's important, choosing what's most important and getting rid of the things that don't really matter. If you do your research, you will find a lot of data in this area. Therefore, you need to narrow down the collected data to select, summarize, and focus on the important data you want.

   **b. Data Presentation**

   After the data is minified, the next step is to look at it. The purpose of presenting the data is to make it easier for the public to understand what is happening and, based on that understanding, plan the next task. Narrative or text is the most common form of data presentation in qualitative research. Report data by explaining the results of the interview to narrative text and attaching supporting files or photo documentation.

   **c. Conclusion Drawing**

   During the research process, that is, during the data collection process, conclusions are drawn continuously. In the research conducted draw conclusions by extracting the essence from a number of sources based on the results of research through observation and interviews and documentation. Conclusions describe relationship patterns, common topics and questions, hypotheses and others.

## III. RESULTS AND DISCUSSION

Information was one of the considerations of investors' decisions. However, not all information that exists, relevant to the interests and intentions of investors. Investors must make quick and informed decisions. ability to achieve expected returns Investment decisions are also closely related to investor behavior. Behavior was the evaluation, feelings, inclination of a person towards something. Such behavior puts a person in the mood to approach and like something or to distance himself and dislike something. In this study, behavior was divided into two areas, namely the perception of behavioral control and the perception of risk. The number of resource persons in this study was 15 informants where these speakers were young investors of Manado State University who were successfully interviewed.

**Table1: Respondents Description**

| No. | Informants | Initial | Investment duration |
|---|---|---|---|
| 1 | Info_1 | JS | 2 Years |
| 2 | Info_2 | Ip | 2 Years |
| 3 | Info_3 | TS | 1 Years |
| 4 | Info_4 | BM | 2 Years |
| 5 | Info_5 | KW | 3 Years |
| 6 | Info_6 | FF | 2 Years |
| 7 | Info_7 | YH | 3 Years |
| 8 | Info_8 | HI | 1 Years |
| 9 | Info_9 | HE | 3 Years |
| 10 | Info_10 | RY | 2 Years |
| 11 | Info_11 | ES | 1 Years |
| 12 | Info_12 | FHS | 3 Years |
| 13 | Info_13 | NY | 3 Years |
| 14 | Info_14 | AR | 2 Years |

From the results of the interview, it can be concluded that investing in the capital market was an alternative that can be done in the midst of this pandemic because it wass difficult to get a job. Furthermore, investment provided profit opportunities due to many companies experiencing a decline in stock prices and the possibility of an increase. We concluded from the answers of respondents from 8 respondents whose answers are almost the same, that before making a decision to buy shares they did an analysis first then they decided on the treatment of a stock to buy it.

Based on the results of interviews, young investors were more confident to buy stocks by analyzing quality of the company's financial statements. The growth of the company also became one of the considerations. Some were doing technical analysis to determine the correct time to buy shares (at a lower prices). The interviews also revealed that informations or advice from friends to buy shares, did not immediately convince them. They tended to re-analyzed the informations. Young investors were considered well literated thus they needed an in-depth analysis of the company instead of blunt advice from other investors.

The next result was the young investors thought that capital was very influential on investment decision making. Less available money to invest resulted in less stocks to buy. The greater the money to invest, the greater the expected profit and vice versa. Based on the results of the interview in making the decision to become an investor, there are several reasons, first because of previous experience then because of the invitation of friends and the prerequisite of Investment Classes.

The results also showed that analysis takes precedence in predicting stock price movements using technical analysis. There are also during the pandemic who prefer to use fundamental analysis because many companies have discounted stock prices so if the analysis was on target, then buy the shares then save them.

*A) Discussion*

From the results of the research that has been described earlier, the author can see that the behavior of individual investors from all aspects of the underlying is very helpful in making their investment decisions. This is based on the following explanation: How investors behave in making investment decisions is viewed from the perception of behavioral control. Based on the results of research when making investment decisions, it shows that the perception of behavioral control was very influential on the decision making of stock investors where confidence and technology have a very clear impact on investor decision making.

All informants in this study stated that before choose the stock to buy, they first analyzed the stocks, from the analysis arose trust, supporting or inhibiting factors based on the investor's behavioral experience. Strong confidence, capital support, and technological support will support the high intention of investors to make a decision such as the decision to choose a stock. Conversely, when investor confidence in self-capital support and technological support is weak, investors' intention to choose stocks is low. The results of this study are in line with Anggraiawan's research (2017) with the title "Determinants of investor behavior in making stock investment decisions in investors registered at GI BEI Telkom University". Based on the results of his research, it shows that the perception of behavioral control has a significant effect on intentions in stock selection.

The discussion above showed the determination of investor behavior in making stock investment decisions at Manado State University. Investors in making investment decisions considering the perception of behavioral control proves that investors at Manado State University are rational investors and are very capable of utilizing technology in their decision making.

One of the requirements for making investment decisions is the availability of information. The problem is that the information available is not all relevant to the interests and goals of each investor. Meanwhile, investors are required to be able to make decisions quickly and precisely. Because, if it is too late or wrong in making decisions, it will result in losing the opportunity to get the expected profit. For this reason, investors need to conduct information analysis in the decision-making process.

Investment decision making was also closely related to investor behavior. Behavior is an evaluation, feeling, tendency of a person towards something. Behavior puts a person on a frame of mind to get closer and like something, or to distance himself and dislike something. Suhari et al. (2011) stated that investor behavior is related to the selection of various investment products and how investors actively act in the capital market.

Private signals dominate investors as a consideration in the investment decision-making process, because investors have a psychological tendency to consider private signals more than public signals. This psychological phenomenon results in the stock price not reflecting its fair price (value). Respondents argue that the Indonesian capital market is in an inefficient condition, managers can choose the right time to issue shares, that is, when the price is high enough above its fair value. That is, the market value tends to be controlled by market participants and does not reflect its fair value.

These personal signals gave rise to investor sentiment. Investor sentiment is the investor's desire to transact based on the company's accounting (fundamental) information. The result of investor sentiment is that investor funds flow into securities that do not provide maximum returns at a certain level of risk. The characteristics of some young investors also determine what factors play an important role in the decision-making process.

Investors should focus on important matters relating to the decisions to be taken. Each investor has the same availability of information. A person's perception will affect him in making decisions. The tendency is that investors will act or move according to their perception.

## IV. CONCLUSION

The investment behavior of young investors at UNIMA in stock investment decision making seen from the perception of behavioral control greatly influences investor decision making that self-confidence through experience and technology has a very clear impact on investor behavior. The results of the interview provide information that before they determine their behavior, they first do a stock analysis, both technical and fundamental analysis, then from the results of the analysis arises trust, supporting factors or obstacles based on the experience of the individual's confession.

**Interest Conflicts**
I hereby declare that there is no conflict of interest concerning the publishing of this paper. This paper purely conducted deliberately on academic purpose.

**Funding Statement**
This paper was self-funded. There is no other party involved in the research and writing process concerning the funding.